\documentclass[prl,twocolumn,showpacs,floatfix]{revtex4}
\usepackage{epsfig}
\usepackage{amsmath,amssymb}
\usepackage{graphicx}
\usepackage{dcolumn}
\usepackage{bm}
\begin{document}
\title{Scaling and Suppression of Anomalous Quantum Decoherence in Ion Traps}
\author{L. Deslauriers}
\author{S. Olmschenk}
\author{D. Stick}
\author{W. K. Hensinger}
\author{J. Sterk}
\author{C. Monroe}
\email{crmonroe@umich.edu}
\affiliation{FOCUS Center and Department of Physics, University of Michigan, Ann Arbor, MI  48109}
\date{\today}
\begin{abstract}
We measure and characterize anomalous motional decoherence of an atomic ion confined in the lowest quantum levels of a novel rf ion trap that features moveable electrodes. The scaling of decoherence rate with electrode proximity is measured, and when the electrodes are cooled from 300 K to 150 K, the decoherence rate is suppressed by an order of magnitude.  This provides direct evidence that anomalous motional decoherence of trapped ions stems from microscopic noisy potentials on the electrodes.  These observations are relevant to quantum information processing schemes using trapped ions or other charge-based systems. 
\end{abstract}
\pacs{32.80.Pj, 39.10.+j, 42.50.Vk}
\maketitle
Trapped atomic ions are an unrivaled source of long-lived entangled quantum states for applications ranging from precision metrology \cite{Squeeze,BeAl} to fundamental studies of quantum nonlocality \cite{Rowe01,Moehring04} and quantum information processing \cite{CZ,NISTJR}. Internal electronic states of multiple ions can be entangled through a mutual coupling to quantum states of the ions' Coulomb-coupled motion, following a number of quantum logic gate schemes \cite{CZ,MS,Milburn}.  These ideas have spawned a flood of recent experimental work, including the generation of particular entangled states of many ions \cite{fourions,eightions,sixions} and the implementation of a variety of simple quantum information algorithms \cite{QDC,FFT,Grover}.

An important source of decoherence in these systems has proven to be the heating of trapped ion motion, thought to arise from noisy electrical potentials on the trap electrode surfaces \cite{NISTJR,turchette,deslauriers04}. This decoherence is expected to become even more debilitating as ion traps become weaker in order to support larger ion crystals \cite{CZ} and allow shuttling of ions through complex and microscale electrode structures \cite{QCCD, SteaneOctupole,RoweQIC,HensingerTee,Stick05}.  More broadly, the electrode surface noise giving rise to anomalous heating in ion traps may also be related to parasitic electrical noise observed in many condensed-matter quantum systems such as Cooper-pair boxes \cite{CPB} and Josephson junctions \cite{JJ}.

In this letter, we present a controlled study of trapped ion motional decoherence in a novel trapping geometry that permits the spacing between the electrodes and the trapped ion to be adjusted \textit{in situ}. The electrodes in this apparatus can also be cooled through contact with a liquid nitrogen reservoir.  When the temperature of the electrodes is reduced from 300 K to about 150 K, the anomalous heating rate of trapped ion motion drops by an order of magnitude or more, but is still higher than that expected from thermal (Johnson) noise in the electrical circuit feeding the electrodes.  These observations provides direct evidence that anomalous heating of trapped ions indeed originates from microscopic ``patch" potentials \cite{patch}, whose fluctuations are thermally driven and can be significantly suppressed by modest cooling of the electrodes.

\begin{figure*}[ptbh]
\centering
\includegraphics[width=1.8\columnwidth,keepaspectratio]{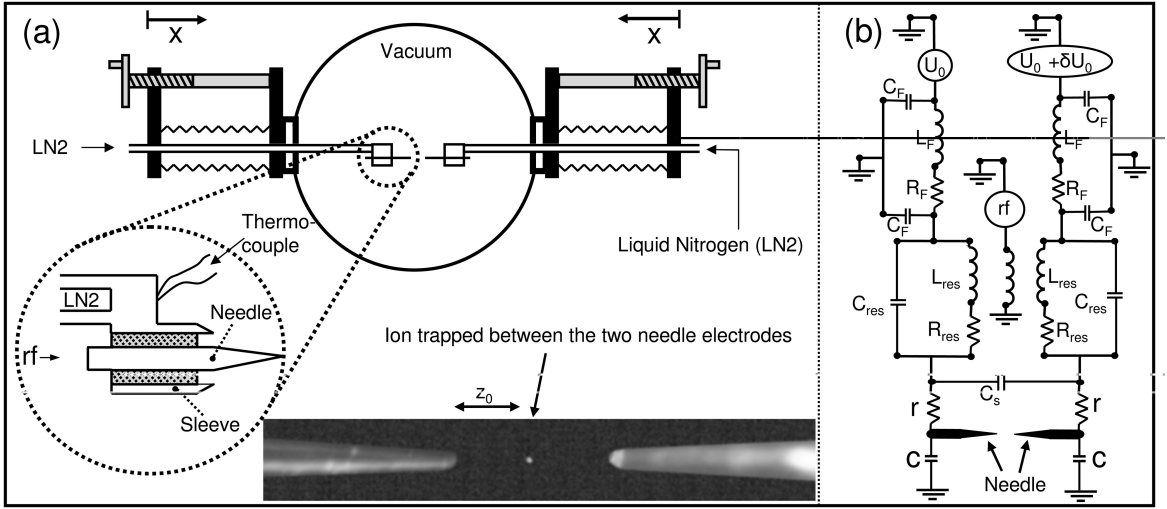}
\caption{(a) Schematic of double-needle rf ion trap.  The needles are separated by variable distance $2z_0$ controlled by external translation stages.  The needle tips are approximately spherical with a radius of $r_0 \approx 3 \mu$m and supported by a conical shank of half-angle $\theta \sim 4^{\circ}$.  Each surrounding cylindrical grounded sleeve of inner diameter $3.0$ mm is recessed $2.3$ mm from the needle tip and electrically isolated from the needles with an alumina tube. (b) Equivalent circuit for the ion trap, including a discrete RLC model of the helical resonators providing rf potentials on each electrode.  The rf resonators are independently biased with static potentials applied through $\pi$-network filters consisting of rf chokes of inductance $L_F = 100 \mu$H and resistance $R_F = 3\Omega$ between $C_F = 0.1 \mu$F capacitors to ground. The outputs of the two rf resonators are shunted with a $C_S = 0.1\mu$F capacitor just outside of the vacuum feedthrough.  The resistance of each electrode lead within the vacuum chamber is approximately $r \sim 0.1\Omega$. Each needle electrode exhibits capacitance $C \sim 5$ pF to its grounded sleeve through the alumina spacer.}
\label{trap}
\end{figure*}

Single cadmium ions are confined in a radiofrequency (rf) trap formed by two opposed needle-tip electrodes in a vacuum chamber, as depicted in Fig. \ref{trap}.  The tungsten electrodes are attached to axial translation stages, allowing the tip-to-tip separation $2z_0$ to be controllably varied over a wide range with micrometer resolution, with coaxial alignment provided by transverse translation stages. A common electrical potential $U_0 + V_0cos(\Omega t)$ with $\Omega/2\pi = 29$ MHz is applied to the electrodes with respect to a pair of recessed grounded sleeves surrounding the needles, as shown in Fig \ref{trap}. The potential is delivered through a dual (bifilar) helical rf resonator, allowing the two electrodes to be independently biased with a static potential difference $\delta U_0$ in order to compensate for any background axial static electric fields.  Ions are loaded into the trap by photoionizing a background vapor of Cd atoms at an estimated pressure of $10^{-11}$ torr.  The average lifetime of an ion in the trap is several hours. 

The axial secular oscillation frequency of an ion in the needle trap is given by 
\begin{equation}
\omega_z = \sqrt{\frac{eU_0\eta}{mz_0^2} + \left(\frac{eV_0\eta}{\sqrt{2}m \Omega z_0^2}\right)^2} 
\label{trapfreq}
\end{equation}
under the pseudopotential approximation ($\omega_z \ll \Omega$) \cite{Dehmelt}, where $e$ is the charge and $m$ the mass of the ion.  The voltage efficiency factor $\eta$ characterizes the reduction in trap confinement compared to the analogous quadrupole rf trap with hyperbolic electrodes of endcap spacing $2z_0$ and ring inner diameter $2\sqrt{2}z_0$ \cite{Dehmelt}.  According to electrostatic simulations of the needle electrodes and grounded sleeves, $\eta = 0.17 \pm 0.01$ over the range of ion-electrode distances $z_0 = 23-250 \mu$m used in the experiment \cite{nosleeve}. The axial secular oscillation frequency of a trapped $^{111}$Cd$^{+}$ ion is measured to be $\omega_z/2\pi = 2.77$ MHz for $U_0=0$ V, $V_0 \sim 600$ V at $z_0 = 136 \mu$m, consistent with simulations.

\begin{figure}[ptbh]
\centering
\includegraphics[width=0.8\columnwidth,keepaspectratio]{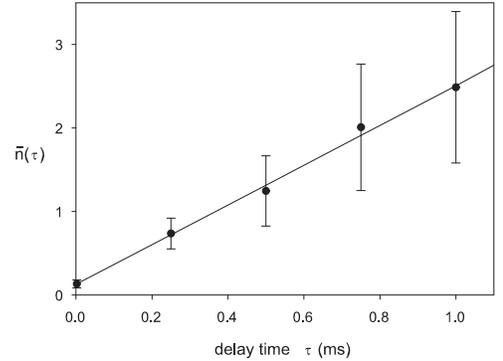}
\caption{Average thermal occupation number $\bar{n}$ measured after various amounts of delay time $\tau$.  The axial trap frequency is $\omega_z/2\pi = 2.07$ MHz with an ion-electrode spacing of $z_0 = 64 \mu$m.  The linear growth in time indicates a motional decoherence or heating rate of $\dot{\bar{n}} = 2380 \pm 440$ quanta/sec.}
\label{nbar}
\end{figure}

In order to measure the decoherence of ion motion in the trap, a single $^{111}$Cd$^+$ ion is first laser-cooled to near the ground state of motion through stimulated-Raman sideband cooling \cite{monroe95,deslauriers04}.  Doppler pre-cooling prepares the ion in a thermal state with an average number of axial vibrational quanta $\bar{n}<20$.  Up to $80$ cycles of pulsed sideband cooling reduce the average occupation number to $\bar{n}<0.3$.  The value of $\bar{n}$ is determined by measuring an asymmetric ratio in the strength of the stimulated-Raman first-order upper and lower sidebands, which is given by $\bar{n}/(1+\bar{n})$ for a thermal state of motion \cite{monroe95, turchette}. We estimate that the systematic error in measuring $\bar{n}$ is no more than $10\%$, from effects such as ion fluorescence baseline drifts, Raman laser intensity imbalances on the sidebands, and non-thermal vibrational distributions.  Motional decoherence is measured by inserting a delay time (up to $\tau=50$ ms) after Raman cooling but before the sideband asymmetry probe. The decoherence or heating rate $\dot{\bar{n}}$ of trapped ion motion is then extracted from the slope of $\bar{n}(\tau)$, as shown in Fig. \ref{nbar}.  
The heating rate of secular motion is related to the power spectrum of electric field noise $S_E(\omega) \equiv \int_{-\infty}^{\infty} \langle E(t) E(t+t')\rangle e^{i\omega t'} dt'$
at the position of the ion by \cite{turchette}
\begin{equation}
\dot {\bar n} = \frac{e^2 }{4m\hbar\omega_z}\left[S_E(\omega_z) + 
\frac{\omega_z^2}{2\Omega^2}S_E(\Omega \pm \omega_z)\right].
\label{heatrate}
\end{equation}
The second term represents the cross-coupling between the noise fields and the rf trapping fields \cite{Dehmelt} to lowest order in the pseudopotential approximation \cite{turchette, NISTJR}.  

\begin{figure}[ptbh]
\centering
\includegraphics[width=0.8\columnwidth,keepaspectratio]{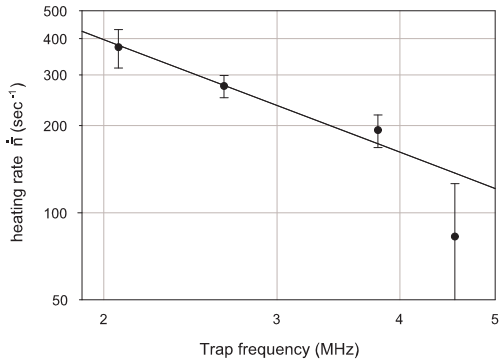}
\caption{Measured axial heating rate $\dot{\bar{n}}$ as a function of axial motional trap frequency $\omega$.  For these data, the ion-electrode spacing is fixed at $z_0 = 103 \mu$m and the trap strength is varied by changing only the static potential $U_0$ while fixing $V_0$, except for the highest frequency point where $V_0$ is $15\%$ higher. The line is a fit to a power law, yielding $\dot{\bar{n}} \sim \omega^{-1.8 \pm 0.4}$, implying that the electric field noise spectral density varies roughly as $S_E(\omega) \sim \omega^{-0.8 \pm 0.4}$ over this frequency range.}
\label{ndot_freq}
\end{figure}

Measurements of ion heating rate are presented in Fig. \ref{ndot_freq} at various trap frequencies $\omega_z$ for a fixed ion-electrode spacing of $z_0 = 103 \mu$m.  In this data set, the trap strength is varied by applying different static potentials $U_0$ while the rf potential amplitude is held constant at $V_0 \approx 600$ V, with an exception for the point at $\omega_z = 4.55$ MHz, where $V_0 \approx 700$ V. The data indicate that the heating rate decreases with trap frequency as $\dot{\bar{n}} \sim \omega_z^{-1.8 \pm 0.4}$, or equivalently that the electric field noise spectrum scales as $S_E(\omega) \sim \omega_z^{-0.8 \pm 0.4}$. Overall, these heating rates are similar to previous measurements in other Cd$^+$ traps of similar size and strength \cite{deslauriers04}, and anomalously higher than the level of heating expected from thermal electric field noise.  

The two needle electrodes are wired to independent but identical circuits (Fig. \ref{trap}b), and the thermal (Johnson) noise from the resistive elements of the electrode circuits can be easily estimated.  A circuit model with discrete elements is justified because the wavelength of the relevant time-varying fields is much larger than the trap structure. 
The net voltage noise across the needle electrodes is 
$S_V(\omega)=8k_BTR(\omega)/[1+R(\omega)^2C^2\omega^2] + 8k_BT'r(\omega)$,
where $k_B$ is Boltzmann's constant.  The first term describes noise originating from a resistance $R(\omega)$ at temperature $T$ attenuated by a capacitive low-pass filter, while the second term describes noise driven directly by resistance $r(\omega)$ at temperature $T'$ after the capactitive filter. The voltage noise $S_V(\omega)$ gives rise to an electric field noise at the ion of $S_E(\omega) = (\epsilon/2z_0)^2 S_V(\omega)$, where the geometrical efficiency factor $\epsilon \approx 0.7$ relates a given potential difference across the needle gap $2z_0$ to the resulting electric field at the ion position.

Thermal noise at frequency $\omega_z$ is driven by the series resistance of the needle electrode tips, estimated to be $\rho/(\pi r_0 tan\theta) \approx 0.1 \Omega$, where $\rho$ is the resistivity of the electrode material at 300 K, $r_0$ is the needle tip radius assumed to be much smaller than the skin depth multipled by $tan\theta$, and $\theta$ is the half-cone angle of the conical shank.  We expect a similar level of thermal noise from the upstream rf choke of resistance $R(\omega_z) = R_F \sim 3 \Omega$, attenuated by filter capacitance $C_F + C_S = 0.2 \mu$F as shown in Fig. \ref{trap}b. The thermal noise at the rf sideband frequencies $\Omega\pm\omega_z$ is dominated by the large effective resistance of the resonator circuit $R(\Omega \pm \omega_z) \approx (\Omega/2\omega_z)^2 R_{res}$ \cite{NISTJR}, where $R_{res} \sim0.1$ $\Omega$ is the dc series resistance of the resonator.  But this near-resonant enhancement is offset by the $(\omega_z/\Omega)^2$ term in Eq. \ref{heatrate}, and because the resulting noise is moreover suppressed by the resonator shunt capacitor $C_S$, this source of thermal noise can be neglected compared to above \cite{rf note}.
In sum, we expect a thermal heating rate of $\dot{\bar{n}} \sim (200/z_0)^2 (\omega_z/2\pi)^{-1}$ sec$^{-1}$ at $300$ K, where $z_0$ is expressed in $\mu$m and $\omega_z/2\pi$ in MHz.  At a temperature of $150$ K, we expect a heating rate a factor of $\sim6$ lower, including a factor of $3$ reduction in the resistivity of the tungsten eletrodes.

\begin{figure}[ptbh]
\centering
\includegraphics[width=1.0\columnwidth,keepaspectratio]{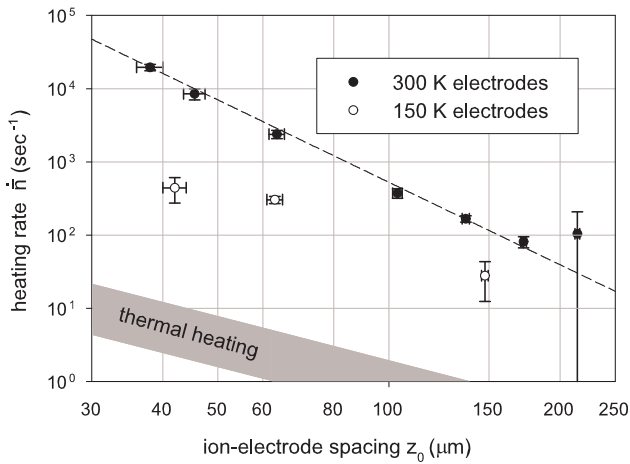}
\caption{Axial heating rate $\dot{\bar{n}}$ as a function of distance $z_0$ from trapped ion to each needle electrode, for warm and cold electrodes. The solid points are measured for 300 K electrodes, and the trap frequency is $\omega_z/2\pi = 2.07$ MHz for this data, with the rf and static trapping potentials increased as the trap is made larger.  The measurements fit well to a scaling of heating rate with trap dimension of $\dot{\bar{n}} \sim z_0^{-3.47\pm0.16}$ (dashed line). The lower set of measurements (open points) are acquired with the needle electrodes cooled to approximately 150 K through contact with a liquid nitrogen reservoir.  These three points were measured on the same trapped ion at a trap frequency of $\omega_z/2\pi = 2.07$ MHz. (The $z_0 = 42 \mu$m cold measurement was performed at $\omega_z/2\pi = 4.9$ MHz and has been scaled upward by a factor of $\sim3$ to the expected heating rate at $\omega_z/2\pi = 2.07$ MHz.) The shaded band at bottom, scaling as $z_0^{-2}$, is the expected range of heating from thermal noise from the trap circuitry.} 
\label{ndot_d}
\end{figure}

Fig. \ref{ndot_d} shows several measurements of heating rates at various values of the distance $z_0$ between the ion and the needle electrodes. This represents the first controlled measurement of heating as a function of electrode proximity, without requiring a comparison across different trap structures, electrode materials, surface qualities, or other factors. In these measurements, the axial trap frequency is held to $\omega_z/2\pi = 2.07$ MHz by varying both the rf and static potentials $V_0$ and $U_0$ as the needle spacing is changed \cite{rfvoltage}.  As seen in the figure, the heating rate fits well to a power law, scaling as $\dot{\bar{n}} \sim z_0^{-3.47 \pm 0.16}$, and is again much higher than that expected from thermal fields. The observed anomalous heating is thought to originate from fluctuating patch potentials on the electrode surfaces, and the scaling of this heating with electrode proximity provides information regarding the spatial size of presumed patches. For a single microscopic patch on the needle electrodes (spatial size much smaller than $z_0$), the electric field noise (e.g., heating rate) is expected to scale as $z_0^{-4}$. For uniformly distributed microscopic patches on the needle electrodes, the electric field noise is expected to scale between $z_0^{-2}$ and $z_0^{-4}$, depending on the details of the electrode geometry. On the other hand, correlated noise across the entire electrode structure (e.g., from thermal Johnson noise or applied voltage noise) is expected to produce a heating rate that scales as $z_0^{-2}$. The observed $z_0^{-3.5}$ scaling rules out correlated noise on the electrodes and is consistent with a model of microscopic patches much smaller than the ion-eletrode spacing $z_0$.  (It is difficult to ascertain the precise average patch size without detailed knowledge of the patch distribution on the electrodes.)

We repeat the measurement of axial motional heating of a trapped ion at various values of $z_0$, but this time with the needle electrodes cooled via contact with a liquid nitrogen reservoir. While the electrode mount is measured to be $80 \pm 5$ K with a vacuum thermocouple, the needle tip is significantly hotter due to absorption of blackbody radiation at 300 K and the limited heat conduction from the very narrow needle electrode to the mount.  We measure the shank of the needle to be $120$ K, and we estimate that the needle tips are cooled to a temperature of $150 \pm 20$ K.  The measured heating rates for cold trap electrodes is plotted in the lower set of data in Fig. \ref{ndot_d}.  These three measurements were performed on one and the same $^{111}$Cd$^+$ ion, over a 6-hour period.  The measured heating rate with cold electrodes is still higher than the expected thermal noise by about two orders of magnitude, as shown in the figure.  However, the nonlinear dependence of observed ion heating with electrode temperature is clear.  The ion heating rate is suppressed by an order of magnitude for a decrease in electrode temperature by only a factor of two, suggesting that the anomalous heating observed in ion traps may be thermally driven and activated at a threshold temperature, and that further cooling to $77$ K or lower may even quench this anomalous heating completely.

In summary, we present a series of controlled measurements of trapped ion motional decoherence, varying both the proximity of the ion to the electrodes and the temperature of the electrodes.  These measurements are performed in a novel moveable ion trap structure that confines laser-cooled ions closer to the electrodes than any previous ion trap.  The measurements of heating in this system provide direct evidence that the heating arises from microscopic fluctuating potentials on the electrodes.  This investigation into a poorly understood source of motional decoherence in trapped ions is of great interest to ion trap quantum information processing and may also have relevance to condensed matter systems that are sensitive to fluctuating electrical potentials.
\begin{acknowledgments}
We acknowledge useful discussions with J. A. Rabchuk and assistance from P. C. Haljan, P. J. Lee and J. Li. 
This work is supported by the National Security Agency and the Disruptive Technology Organization under Army Research Office contract W911NF-04-1-0234, and the National Science Foundation ITR Program.
\end{acknowledgments}
\bibliographystyle{prsty}

\begin{thebibliography}{10}
\bibitem{nosleeve} Without the grounded sleeves, simulations predict that the voltage efficiency factor drops to $\eta \sim 0.06$, as the effective ground electrode is much further the needle tips.  This is consistent with measurements of the trap frequency in an earlier (sleeveless) version of the trap.
\bibitem{rf note} For ``symmetric" trap designs where the rf electrodes symmetrically surround the ion and are wired together, the resulting common-mode thermal noise from the rf leads should vanish due to symmetry \cite{Stick05}.
\bibitem{rfvoltage} Previous measurements of axial heating of trapped Cd$^+$ ions in a linear trap \cite{deslauriers04} show no correlation between the heating rate and the rf trapping voltage amplitude $V_0$.  These earlier measurements were performed over similar ranges of rf voltages and trap frequencies as used for the current experiment.  We therefore assume that the origin of anomalous heating measured for the needle trap is not directly influenced by the rf voltage amplitude.
\end{thebibliography}

\end{document}